\documentclass[12pt]{article}
\begin{document}
\title{Comment on the paper "Weyssenhoff fluid dynamics
in a 1+3 covariant approach" (arXiv:0706.2367)}
\author{D. PALLE \\
Zavod za teorijsku fiziku \\
Institut Rugjer Bo\v skovi\'c \\
Po\v st. Pret. 180, HR-10002 Zagreb, CROATIA}
\date{ }
\maketitle

\vspace{5 mm}
{\it
Few comments are given to clarify some issues of Weyssenhoff
fluid in the Einstein-Cartan gravity.
}
\vspace{5 mm}

The authors of the paper "Weyssenhoff fluid dynamics ..." \cite{Brechet} Brechet, Hobson and Lasenby compare their treatment and
results with my paper "On primordial cosmological density ..."
\cite{Palle}. I list some clarifications and comments:

(1) The relation (4) from \cite{Palle} is just the
generalized Ricci identity derived by Hehl (see Eq. (3.41)
of \cite{Hehl}). "Effective field equations" are valid
generally in Riemann-Cartan spacetimes, and not only for the Weyssenhoff fluid. This is also proved by Hehl (see Eq. (3.78)
of \cite{Hehl}). It is just the consequence of the 
algebraic relation between spin and torsion (namely, the
absence of spacetime derivatives).

(2) Conformal Weyl tensor does not appear in field equations,
but only in Ricci and Bianchi identities describing tidal forces
in clumpy Universe. However, the Universe is clumpy at small scales from dark ages
to the present. The N-body cosmic simulations with
Newtonian gravity imbedded into some background geometry (homogeneous or
inhomogeneous, isotropic or anisotropic) are the most suitable 
theoretical tools in such a situation, certainly not the Weyssenhoff fluid in general relativity.
This is the reason why I do not include Weyl tensor at large 
scale and large redshift considerations.

(3) The evolution equations are derived in \cite{Palle}
by combining Ricci identities and field equations.
The conservation equations are similarly obtained from Bianchi identites
and field equations by Obukhov and Korotki \cite{Obukhov}.
It suffices then for the main topic of my paper to 
study the evolution of the mass-density contrast 
at large scales and large redshifts within the
gauge invariant formalism of Ellis and Bruni \cite{Ellis}.

\end{document}